\documentclass[preprint,authoryear]{elsarticle}

\usepackage{graphicx}

\usepackage{amssymb}

\journal{Journal of Theoretical Biology}

\begin{document}

\begin{frontmatter}


%
\title{Interference competition and invasion: spatial structure, novel weapons and resistance zones}
%
%
\author[suny]{Andrew Allstadt\fnref{fn1}}
\ead{andrew.allstadt@virginia.edu}
\author[suny]{Thomas Caraco\corref{cor1}}
\ead{caraco@albany.edu}
\author[rpi]{F. Moln\'ar Jr.}
\ead{molnaf@rpi.edu}
\author[rpi]{G. Korniss}
\ead{korniss@rpi.edu}
\fntext[fn1]{Present address: Blandy Experimental Farm, University of Virginia, Boyce, VA 22620, USA}
\cortext[cor1]{Corresponding author.  Tel.: +1 518 442 4343}
\address[suny]{Department of Biological Sciences, University at Albany, Albany, NY 12222, USA}
\address[rpi]{Department of Physics, Applied Physics and Astronomy, Rensselaer Polytechnic Institute, Troy, NY 12180, USA}
\begin{abstract}
Certain invasive plants may rely on interference mechanisms (allelopathy, \emph{e.g}.) to gain competitive superiority over native species.  But expending resources on interference presumably exacts a cost in another life-history trait, so that the significance of interference competition for invasion ecology remains uncertain.  We model ecological invasion when combined effects of preemptive and interference competition govern interactions at the neighborhood scale.  We consider three cases. Under ``novel weapons,'' only the initially rare invader exercises interference.  For ``resistance zones'' only the resident species interferes, and finally we take both species as interference competitors.  Interference increases the other species' mortality, opening space for colonization.  However, a species exercising greater interference has reduced propagation, which can hinder its colonization of open sites.  Interference never enhances a rare invader's growth in the homogeneously mixing approximation to our model.  But interference can significantly increase an invader's competitiveness, and its growth when rare, if interactions are structured spatially.  That is, interference can increase an invader's success when colonization of open sites depends on local, rather than global, species densities.  In contrast, interference enhances the common, resident species' resistance to invasion independently of spatial structure, unless the propagation-cost is too great.  Increases in background mortality (\emph{i.e}., mortality not due to interference) always reduce the effectiveness of interference competition.
\end{abstract}

\end{frontmatter}


\section{Introduction}
Both lateral and vertical interactions can affect the likelihood that an invasive species advances when rare \citep{Levine_2004,Going_2009}.  But for many plants, competitive asymmetry between invader and resident species governs both the outcome and the timescale of ecological invasion \citep{Molofsky_2004,Vila_2004,OMalleyTPB_2006,MacDougall_2009}.  Given invader-resident competition, ecological superiority may depend on more than one mechanism \citep{Case_1974,Ridenour_2001}.
Our analysis addresses the combined impact of preemptive and interference competition on invasion dynamics when biotic interactions are structured spatially.  We assume that interference has a cost \citep{Adams_1979,Bazzaz_1997,Amarasekare_2002};
an increasing level of interspecific interference requires a reduction in propagation rate, diminishing that species' capacity to colonize unoccupied sites.

In many plant communities the primary mode of competition
is site preemption \citep{Schoener_1983,Bergelson_1990,Crawley_1999,Yurkonis_2004}; \emph{i.e}., species interact through colonization of empty sites.  Superior preemptive competitors have higher propagation rates or lower mortality rates
\citep{Korniss_JTB2005,OMalley_PRE2006,Allstadt_2007}.
The former increases colonization of open sites, and the latter decreases a competitor's opportunities for colonization.  Preemptive competitors have the same niche in a spatially homogeneous environment \citep{Amarasekare_2003}.  Ordinarily this precludes coexistence \citep{Shurin_2004,Allstadt_2009}, since self-regulation does not exceed interspecific competition.
\citet{Case_1974} suggest that this niche similarity might favor evolution of interference mechanisms.

An interference competitor inhibits another species' access to a critical resource, often by harming individuals of the other species.  Examples include interspecific territoriality in animals and chemical competition in plants
\citep{Case_1994,Callaway_2000}.  Exotic invaders may suppress native species' densities through interference competition \citep{DAntonio_1992,Callaway_2004,Cappuccino_2006}.
The ``novel weapons'' hypothesis proposes that some invasive plants release chemicals that inhibit growth of native species \citep{Callaway_2000}.  Interestingly, allelopathic interference may act directly on individuals of the resident competitor, or may act indirectly through toxic effects on native species' microbial mutualists, particularly mycorrhizal fungi
\citep{Wolfe_2005,Callaway_2008}.  So, under the novel weapons hypothesis, invaders attain competitive superiority because their phytochemicals present novel challenges to native species.  Reasonably, in other communities, exotic species likely encounter interference competition from natives; see comments in \citet{vHolle_2003}.  Our models address the three general cases where interference can affect the outcome of resident-invader competition.  We associate novel weapons with interference by the invader only.  We refer to ``resistance zones'' when the resident species, but not the invader, exerts interference competition.  And, of course, invaders and residents may each compete \emph{via} interference \citep{Case_1974}.  If neither species exhibits interference, our model leaves preemption as the sole competitive mechanism.  In this case the species that, when alone, maintains the greater equilibrium density will always displace its competitor \citep{Amarasekare_2003,OMalleyTPB_2006,Allstadt_2007}.

Discrete, stochastic spatial models and their deterministic
analogues have been employed, commonly and successfully, to gain
insight into collective behavior of multi-``species'' interactions
\citep{Marro_1999,Murray_2003} in physics
\citep{Korniss_EPL1995,Korniss_JSP1997}, chemistry
\citep{Antal_JCP1998,Toro_EPL1997,Ziff_PRL1986}, and in the study of
population dynamics
\citep{Ellner_1998,McKane_PRE2004,OMalley_PRE2006,OMalley_BMAB}. We
use both methods to investigate the combined effects of preemptive
and interference competition on invasion. We organize our paper as
follows.  First, we present a discrete (individual-based),
stochastic model where an invader and a resident species compete
preemptively, and one or both species also employs interference.  We
let a species' propagation rate depend functionally on its level of
interference; we consider convex, linear and concave trade-offs.  We
explore the model by analyzing invasibility criteria of both a
mean-field approximation (homogeneous mixing) and a pair
approximation.  Then we apply results of the approximations to
interpret simulations of the full spatial model.  The Discussion
compares our results to other spatial models incorporating
allelopathy.  Appendices collect much of the analytical detail.

Our results find that interference by the invader increases its likelihood of successful invasion only when interactions are spatially structured.  That is, the novel weapons advantage, in our model, appears only when we account for local clustering of invader individuals.  Interference by the resident, the initially common species, can inhibit invasion with and without spatial structure.  Increases in background mortality rate (mortality before adjustment due to interference) diminish the competitiveness of the species relying more on interference.  Increased mortality reduces both local and global densities; that is, the frequency of empty sites increases.  Consequently, the value of interference, relative to propagation, declines.  Our model assumes that interference increases mortality of a preemptive competitor, but interference does not generate an alternative niche.  Consequently, we should not anticipate coexistence absent continuous introduction or strong effects of spatial clustering \citep{Allstadt_2009}.

\section{A discrete, stochastic spatial model}
When plant species compete, interactions regulating population growth generally occur at the neighborhood scale \citep{Goldberg_1987,Uriarte_2004}.  Interference competition, including allelopathy, ordinarily has a local spatial structure.  And, when propagule dispersal distance (inter-ramet distance in clonal species) is limited, plants compete preemptively for space at the neighborhood scale.  Consequently, our model - which integrates preemptive and interference competition - assumes that competitive interactions among nearest neighbors drive invasion and the dynamics of species' abundances \citep{OMalleyTPB_2006}.
\subsection{Model construction}
Two clonal plant species compete on an $L_x$ $\times$ $L_y$ lattice with periodic boundaries.  Each lattice site represents the resources required to sustain a single individual (a ramet) of either species.  The local occupation number at site ${\bf x}$ is $n_i({\bf{x}})=0,1$ with $i= 1,2$, referring to the resident and invader species, respectively.  During a single simulated time unit, one Monte Carlo step per site [MCSS], $L_xL_y$ sites are chosen randomly for updating.

An empty site may be occupied by species $i$ through introduction from outside the environment, or through local propagation.  Introduction of species $i$ at an open site occurs as a Poisson process with rate $\beta$.  Each species has the same introduction rate, to avert any effect of propagule pressure on the outcome of competition. Local propagation into an open site has rate $\alpha_i\eta_i({\bf x})$, where $\alpha_i$ is the individual-level propagation rate for species $i$, and
$\eta_i({\bf x}) = (1/\delta)\Sigma_{{\bf x}'\epsilon {\rm nn}({\bf x})}n_i({\bf x}')$ is the density of species $i$ in the neighborhood around open site ${\bf x}$.  ${\rm {nn}}({\bf x})$ is the set of nearest neighbors of site ${\bf x}$, and $\delta$ is the number of sites in that neighborhood ($\delta=|nn({\bf x})|$).  Since we equate colonization with propagation of new ramets, we let $\delta=4$.

Colonization can occur only at open sites.  An occupied site opens through mortality.  Individuals of each species suffer density-independent mortality at rate $\mu$ \citep{Cain_1995}.  An individual occupying site $\textrm{\textbf{x}}$ experiences an increased mortality rate due to interference if $nn({\bf x})$ includes any heterospecifics.  That is, an individual of species $i$ at site $\textrm{\textbf{x}}$ has total mortality rate $\mu + \theta_j \eta_j(\textrm{\textbf{x}}),~i\neq j$.  $\theta_j \geq 0$ represents interference by species $j$, and $\eta_j(\bf x)$ is the density of species $j$ on the neighborhood around site $\textrm{\textbf{x}}$.

Summarizing transition rules for an arbitrary site ${\bf x}$, we have
\begin{equation}
0\stackrel{\beta + \alpha_1\eta_1(\bf{x})}{\longrightarrow}1, \;\;
0\stackrel{\beta + \alpha_2\eta_2(\bf{x})}{\longrightarrow}2, \;\;
1\stackrel{\mu + \theta_2 \eta_2 (\bf {x})}{\longrightarrow}0, \;\;
2\stackrel{\mu + \theta_1 \eta_1 (\bf{x})}{\longrightarrow}0, \;\;
\label{localrates}
\end{equation}
where 0, 1, 2 indicates whether a site is open, resident-occupied, or invader-occupied, respectively.  Table~\ref{table_params} defines the symbols and notation we use.
\begin{table}[t]
\caption{Definitions of model variables and parameters \label{table_params}}
\begin{tabular}{|c|l|}
  \hline
  Symbols & Definitions \\
  $L_x ,\; L_y ( = L)$ & Lattice size \\
  {\bf x} & Location of lattice site \\
  $n_1({\bf x})$ & Occupation number for residents at site ${\bf x}$; $n_1({\bf x})= 0,1$ \\
  $n_2({\bf x})$ & Occupation number for invaders at site ${\bf x}$; $n_2({\bf x})= 0,1$ \\
  nn({\bf x}) & Set of nearest neighbors around site \textbf{x} \\
  $\delta$ & size of neighborhood around site ${\bf x}$ ($\delta=|nn({\bf x})|$) \\
  $\beta$ & Common introduction rate at empty sites \\
  $\eta_i({\bf x})$ & Density of species $i$ on nn(\textbf{x}) \\
  $\alpha_i$ & Individual rate of propagule production, species $i$ \\
  $\mu$ & Background mortality rate, both species \\
  $\theta_j$ & Mortality of species $i$ due to interference by species $j$ \\
  $\alpha_c (\mu)$ & Minimal propagation rate for persistence \\
  $C$ & Maximal propagation rate; C = 0.8 \\
  $R$ & Curvature of $\alpha , \theta$ trade-off \\
  ${\rho_i}^*$ & Equilibrium single-species density \\
  \hline
\end{tabular}
\end{table}
We assume that interspecific competition drives the dynamics; \emph{i.e}., each species persists absent competition.  Therefore, we restrict attention to the
$\beta \ll \alpha_c(\mu) < \alpha_i$ $(i = 1, 2)$ regime, where $\alpha_c(\mu)$ is the critical propagation rate below which either species, in the other's absence, grows too slowly to avoid extinction \citep{Oborny_2005,OMalleyTPB_2006}.
\subsection{Life-history constraint}
The essential lesson of life-history theory is that increased allocation of resources to one trait advancing survival or reproduction comes at the expense of another trait \citep{Bell_1986}.  Applying this concept, we assume that any increase in a species' level of interference reduces that species' clonal propagation rate.  The functional dependence (trade-off) has the form $\alpha_i^R + \theta_i^R = C^R$, where $C$ is a constant, equal to the maximal feasible rate of propagation, and $R$ defines the shape of the trade-off \citep{GandC_2000}.  If $0 < R < 1$, the cost of increasing $\alpha$ (i.e., the decrease in $\theta$) decreases as $\alpha$ increases.  If $R = 1$, the cost of increasing $\alpha$ is constant, and if $R > 1$, the cost of increasing $\alpha$ increases at greater $\alpha$.  More importantly, any increase in $R$ increases competitiveness of a species that employs \emph{both} preemption and interference; the constraint moves farther from the origin as $R$ increases.  If a species does not exercise interference, we let its propagation rate vary on ($\alpha_{c}(\mu)$, 0.8].  Therefore, we let $C = 0.8$.
\subsection{Novel weapons}
Under the novel weapons hypothesis, the invader exercises interference competition while the resident does not.  That is, $\theta_1 = 0$, and
\begin{displaymath}
0 < \theta_2 \leq \sqrt[R]{C^R - {\alpha_c(\mu)}^R}
\end{displaymath}
Invasive plants may interfere with native competitors through several mechanisms.  Some impact natives directly; others are mediated through a third species.  Invaders can act as a disease reservoir \citep{Eppinga_2006,Borer_2007}, focus herbivory on native species \citep{Dangremond_2010}, or produce allelopathic chemicals \citep{Prati_2004,Cippolini_08}.  Our analysis emphasizes the invader's trade-off between propagation and interference across a range of native species lacking interference.  Under the novel weapons hypothesis, the resident species does not compete through interference, and we let its reproductive rate vary to model variation in the strength of preemption the invader encounters.
\subsection{Resistance zones}
To study resistance zones, we let the resident exercise interference competition while the invader does not.  That is, $0 < \theta_1 \leq \sqrt[R]{C^R - {\alpha_c(\mu)}^R}$, and $\theta_2 = 0$.  A resident species exercising interference competition may strongly resist invasion since its initial density (by definition) will be relatively high.  Given biotic resistance combining preemption and interference, we vary the invader's propagation rate to consider a range of matches between the invader and the abiotic environment invaded.
\subsection{Mutual interference}
Under the assumption of mutual interference competition, we have $0 < \theta_1 , \theta_2 \leq \sqrt[R]{C^R - {\alpha_c(\mu)}^R}$.  For this case we assume that each species' resource allocation between propagation and interference follows the same trade-off.
\section{Mean-field approximation}
The spatially-homogeneous mean-field approximation to our model ignores any effects of locally clustered growth.  The mean-field model (MF) so offers comparison to results including interactions at the neighborhood scale; see \citet{Wilson_1998}, \citet{Pascual_1999} or \citet{Cuddington_2000} for perspective.

$\rho_1$ and $\rho_2$ represent the global densities of species 1 and species 2, respectively.  Ignoring continuous introduction (letting $\beta = 0$) we have a mean-field dynamics:
\begin{eqnarray}
\dot{\rho_1} & = & \alpha_1 \rho_1 (1 - \rho_1 - \rho_2) - \rho_1 (\mu + \theta_2 \rho_2 )\\
\dot{\rho_2} & = & \alpha_2 \rho_2 (1 - \rho_1 - \rho_2) - \rho_2 (\mu + \theta_1 \rho_1 )
\label{InterfereMF}
\end{eqnarray}
$\theta_i$ represents
the increased mortality of species $j$ induced by species $i$; we scale $\theta_i$ per unit density of species $i$.
Species' persistence absent competition in the MF approximation requires only that each $\alpha_i > \alpha_c(\mu) =\mu$.  Letting $\theta_i \geq 0$, for $i = 1,~2$, we conduct a general stability analysis of the MF model's equilibria.  Thereafter, we consider competitive superiority when the specific propagation-interference constraint introduced above applies.
\subsection{Equilibria and stability}
The dynamics has three boundary equilibria and one positive equilibrium.  Mutual extinction, designated equilibrium E1, cannot be stable since each $\alpha_i > \mu$.  At Equilibrium $E2$, species 1 competitively excludes species 2:
\begin{equation}
E2: ~\left( {\rho_1}^* = 1 - \frac{\mu}{\alpha_1}, {\rho_2}^* = 0\right)
\label{E2}
\end{equation}
Species 2 excludes species 1 at equilibrium $E3$:
\begin{equation}
E3: ~\left( {\rho_1}^* = 0, {\rho_2}^* = 1 - \frac{\mu}{\alpha_2}\right)
\end{equation}
Finally, a positive (internal) equilibrium, $E4$, can be expressed as:
\begin{eqnarray}
{\rho_1}^* = \frac{\theta_2 (\alpha_2 - \mu ) + \mu (\alpha_2 - \alpha_1)}{\alpha_1 \theta_1 + \alpha_2 \theta_2 + \theta_1 \theta_2}\\
{\rho_2}^* = \frac{\theta_1 (\alpha_1 - \mu ) + \mu (\alpha_1 - \alpha_2)}{\alpha_1 \theta_1 + \alpha_2 \theta_2 + \theta_1 \theta_2}
\label{E4}
\end{eqnarray}
As noted above, we know that when competition is strictly preemptive $(\theta_1 = \theta_2 = 0)$, the species cannot coexist under homogeneous mixing.

Suppose that the resident excludes the invader.  From Appendix A, local stability of invader extinction requires:
\begin{equation}
\mu (\frac{\alpha_2}{\alpha_1} - 1) + \theta_1 (\frac{\mu}{\alpha_1} - 1) < 0
\label{StabE2}
\end{equation}
Since $\mu < \alpha_1$, $\theta_1 (\frac{\mu}{\alpha_1} - 1) \leq 0$.
Therefore, any interference by the resident tends to stabilize invader extinction by increasing invader mortality.  That is, interference by the common species can help prevent advance of the rare species, even if the rare species has the greater propagation rate.  Clearly, if
$\alpha_1 > \alpha_2$ and $\theta_1 \geq 0$, the resident species repels the invader.  Note that $\theta_2$, the invader's level of interference competition, does not appear in Inequality (\ref{StabE2}).  Hence the invader cannot increase its growth rate when rare through interference; only increased propagation promotes mean-field invasion.

To elaborate, the invader advances from rarity only if Inequality (\ref{StabE2}) is reversed.  From Appendix A, assuming $\theta_1 > 0$, the rare species invades \emph{iff}:
\begin{equation}
\alpha_2 - \alpha_1 > \theta_1 (\frac{\alpha_1}{\mu} - 1) > 0
\label{2invades}
\end{equation}
To invade successfully, the rare species must have the greater propagation rate, and must not suffer too much interference from the common species.

Appendix A shows that if species 2 can invade the resident, then equilibrium $E3$ is locally stable.  That is, if species 2 can advance when rare, it will grow to exclude species 1.

\subsection{Bistability}
Bistability requires that local-stability conditions for  both single-species equilibrium nodes ($E2$ and $E3$) hold simultaneously.  Given bistability, initial conditions determine the outcome; at some point, the more abundant species ``wins.''  From Appendix A, the common species (1 or 2) repels its rare competitor if
\begin{equation}
\theta_1 (1 - \frac{\alpha_1}{\mu}) < \alpha_1 - \alpha_2 ~~\textrm{and}~~\theta_2 (\frac{\alpha_2}{\mu} - 1) > \alpha_1 - \alpha_2
\label{bistable}
\end{equation}
If $\alpha_1 > \alpha_2$, the first expression must hold, since $\theta_1 (1 - \frac{\alpha_1}{\mu}) \leq 0$.
The second inequality can hold simultaneously if $\theta_2$ is relatively large, and the difference between
propagation rates is not too large.  If $\alpha_2 > \alpha_1$, the second expression must hold, since $\theta_2 (\frac{\alpha_2}{\mu} - 1) > 0$, and symmetric conditions promoting bistability are clear.  Bistability can arise even if the species with the greater propagation rate does not exert interference.  For example, if $\alpha_1 > \alpha_2$, the system can be bistable when
$\theta_1$ = 0 (provided, of course, $\theta_2 > 0$).

\begin{figure}[b]
\centering
\vspace*{5.40truecm}
       \includegraphics{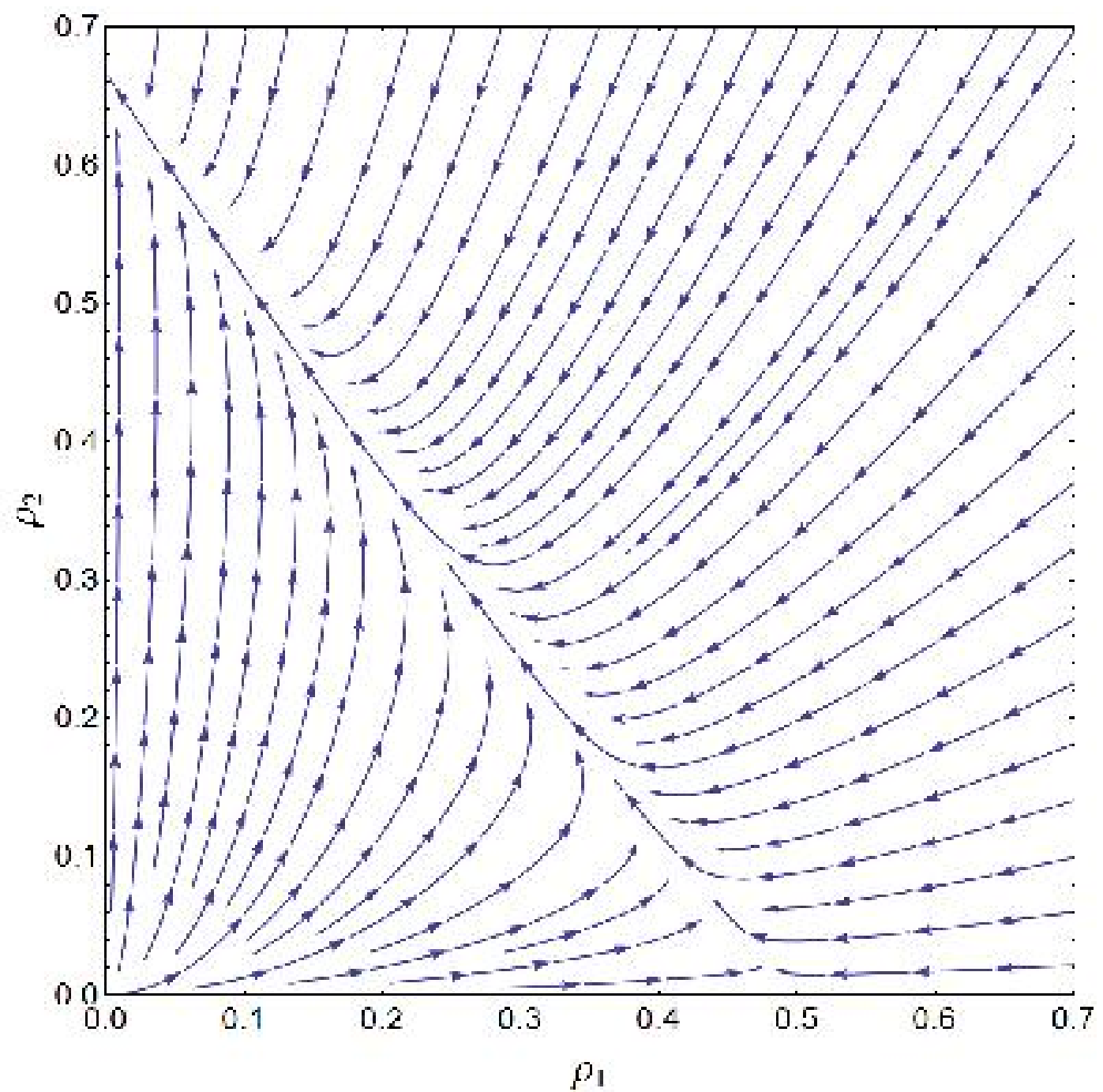}
\vspace*{5.40truecm}
       \includegraphics{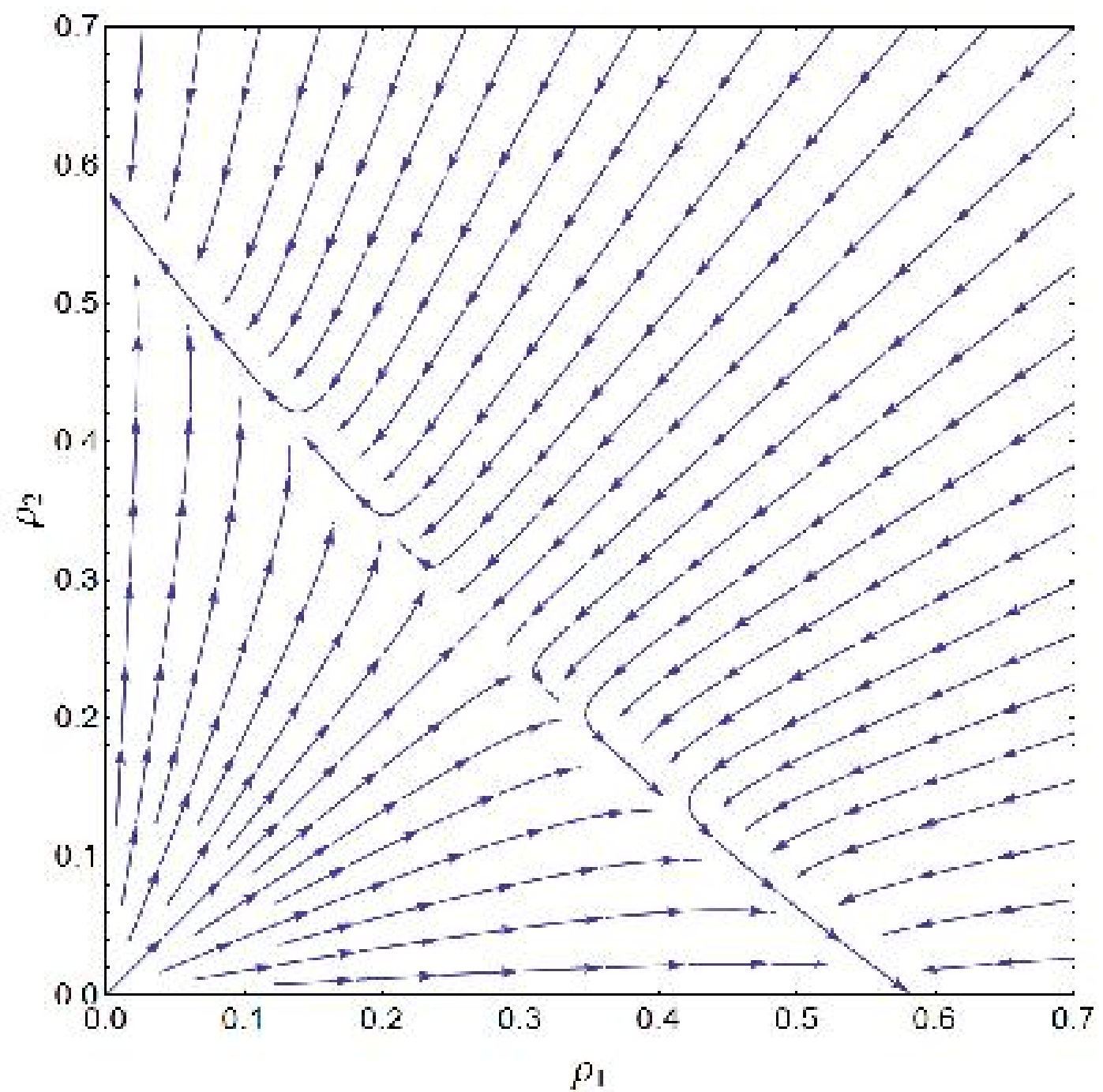}
\vspace*{5.40truecm}
       \includegraphics{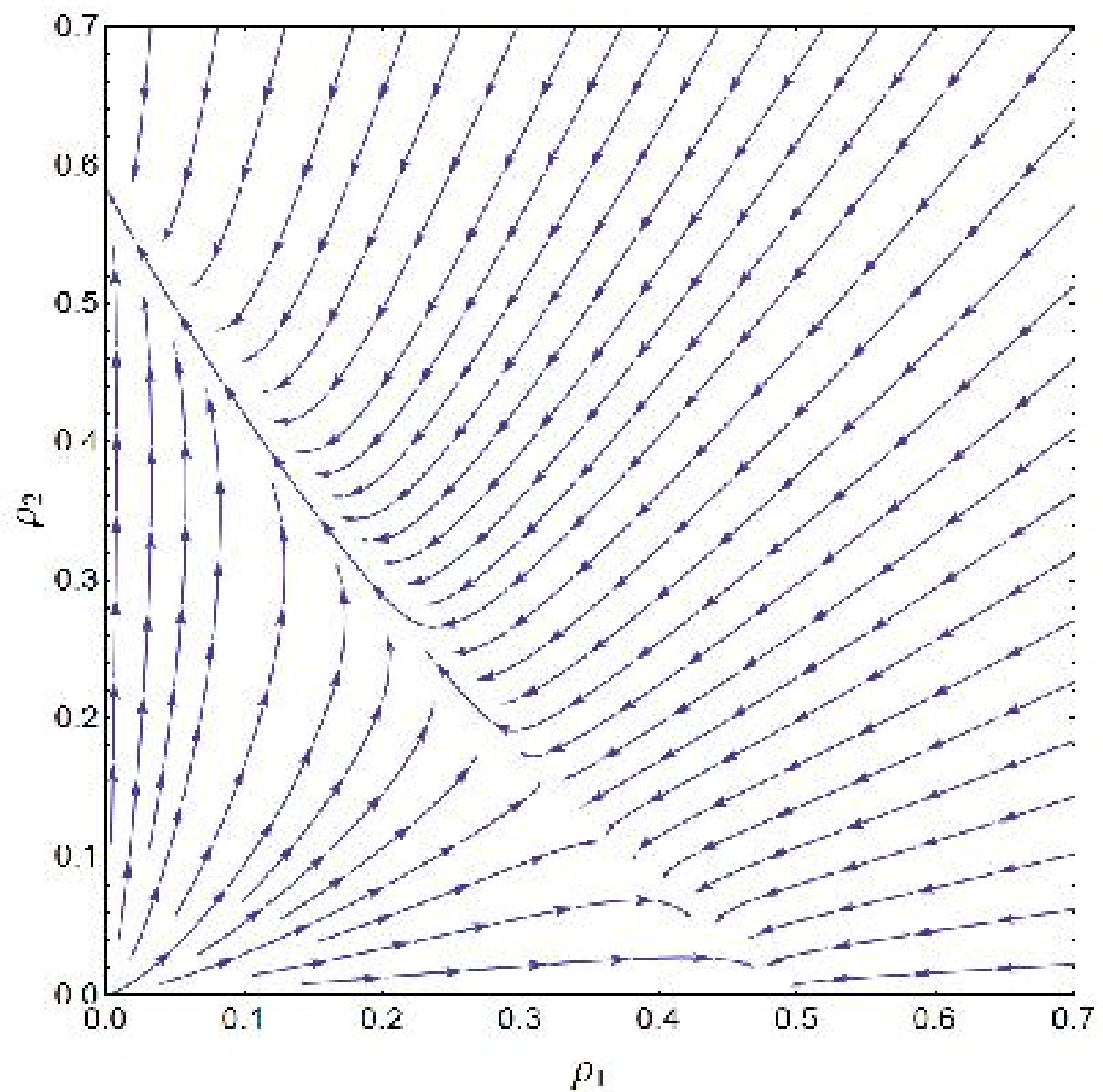}
\caption{
Flow in the mean-field dynamics over the $(\rho_1, \rho_2)$ phase space.  $\mu = 0.25$ in each plot.  Top: Species 2 excludes species 1 independently of initial condition; Expression (\ref{2invades}) holds.  $\alpha_1 = 0.5$, $\theta_1 = 0.2$, and $\alpha_2 = 0.75$. Middle: Condition for bistability, Expression (\ref{bistable}), holds.  $\alpha_1 = 0.5 = \alpha_2 = 0.6$, $\theta_1 = \theta_2 = 0.1$.  Bottom: Bistable dynamics; $\alpha_1 = 0.5 = \alpha_2 = 0.6$, $\theta_1 = 0.2, \theta_2 = 0.1$.  Species 2 has the greater domain of attraction.}
\label{flowplots}
\end{figure}

Figure \ref{flowplots} shows how the mean-field dynamics can flow in in the $(\rho_1, \rho_2)$ phase space.  One example plots competitive asymmetry.  Species 2 advances when rare, and when common, excludes species 1.  The other plots show bistability, one for a symmetric (identical species), and one for an asymmetric, domain of attraction.
\subsection{Coexistence?}
Coexistence (local stability of a positive equilibrium) requires that each species invade the other.  Together, the conditions for mutual invasion indicate that coexistence requires (Appendix A):
\begin{equation}
\theta_1 (1 - \frac{\alpha_1}{\mu}) > \alpha_1 - \alpha_2~~\textrm{and}~~\theta_2 (1 - \frac{\alpha_2}{\mu}) > \alpha_2 - \alpha_1
\label{coexist}
\end{equation}
Since each $\alpha_i > \mu$, each $(1 - \frac{\alpha_i}{\mu}) < 0$; the LHS of each of these inequalities must then be non-positive.  But one $(\alpha_i - \alpha_j)$ must be positive, so that the two inequalities cannot be true simultaneously.  Hence the MF does not permit mutual invasion, and so does not admit competitive coexistence, when the species interact through both  preemption and interference.
\subsection{Propagation-interference constraint}
\begin{figure}[b]
\centering
\vspace*{6.00truecm}
       \includegraphics{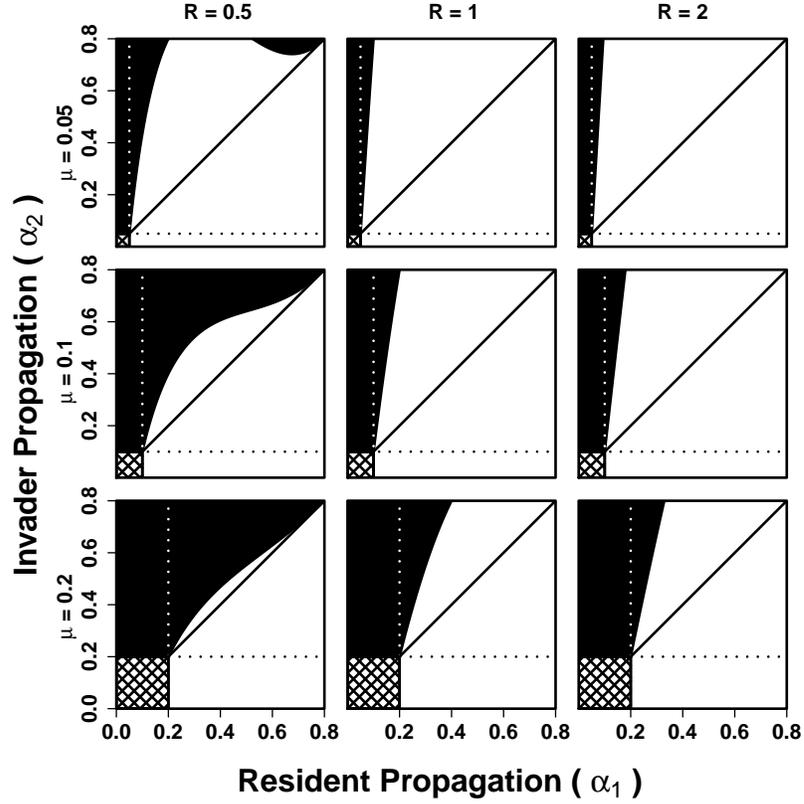}
\vspace*{4.0truecm}
\caption{
Pairwise invasion plots generated by the mean-field approximation.  Each subplot's ordinate is the resident's propagation rate, $\alpha_1$.  The resident's level of interference is given by the life-history constraint, $\alpha_1^R + \theta_1^R = C^R$; The MF invasion criterion is insensitive to interference by the invader.  Black: parameter combinations where invader's growth is positive when rare; invasion succeeds. White: resident repels invader; invasion fails. Separation of these equilibrium phases follows from Inequality (\ref{invoke1}).  Note that the diagonal (solid line) separates invader success from failure when each species competes through site preemption only.  Rows, top to bottom, show that invasion increases as background mortality increases.  Columns, left to right, show that invader's success declines as resident propagation increases for given level of interference.  If any of the plots is reflected along the diagonal, parameter combinations where ``white falls on white'' fulfill criteria for bistability.  Dotted lines denote $\alpha_c(\mu)$ for each species, and neither species is viable in the hatched region.
}
\label{MF}
\end{figure}

The preceding, general stability analyses did not treat interference as a function of  propagation rate.  Invoking the life-history constraint generates a special case of the MF  analysis.  The condition for the invader's advance when rare, Inequality (\ref{2invades}), becomes:
\begin{equation}
\alpha_2 > \alpha_1 + \Big(\frac{\alpha_1}{\mu} - 1\Big) \sqrt[R]{C^R - {\alpha_1}^R}
\label{invoke1}
\end{equation}
From the general MF analysis, any allocation to interference by the invader diminishes its propagation rate $(\alpha_2)$, and so decreases the likelihood of successful invasion.  Any increase in $R$, relaxing the propagation-interference tradeoff, increases the range of feasible parameter combinations where the resident repels the invader (see fig. \ref{MF}).  That is, the opportunity for successful invasion declines as $R$ increases in the MF model, since only the resident can benefit from interference under homogeneous mixing.

Increasing background mortality $(\mu)$ increases the likelihood of successful invasion.  Greater mortality, with $\theta_1$ and the $\alpha_i$ fixed, decreases the resident's density.  Consequently, sites available for colonization increase in density, and the impact of interference on the invader's dynamics declines.  Therefore, greater background mortality, which affects both species, decreases the range of feasible parameter combinations where the resident repels the invader (see fig. \ref{MF}).

If the competitors mix homogeneously, interference can help the resident repel the invader, but cannot promote the rare species' invasion.  Interference, as a mechanism of biotic resistance, has a greater effect when a small level of interference does not cost too much in propagation, and when background mortality is relatively low (so that resident density is relatively high).
\section{Pair approximation}
Pair approximation (PA) incorporates correlations of the occupation status of nearest neighboring sites into a dynamics \citep{Dickman_1986,Matsuda_1987,Bauch_2000,vBaalen_2000}.  By incorporating a minimal spatial structure, PA predicts equilibrium densities of individual-based models more accurately than do MF models \citep{Ellner_1998,Caraco_2001}.  For our purposes, PA addresses consequences of local clustering, generated by dispersal limitation, for invasion.  Appendix B presents details.

The PA tracks global densities $\rho_i(t)$, and conditional densities $q_{j|i}(t)$ representing the likelihood that a site is in state ${j}$, given that a neighboring site in state ${i}$ \citep{Sato_2000,OMalleyTPB_2006}.  The three global densities ($\rho_0(t), \rho_1(t), \rho_2(t)$) imply 9 local densities (${q_{j|i}}$). However, the PA's dimension is limited by simple constraints:
\begin{eqnarray}
\rho_0 + \rho_1 + \rho_2 = 1 \nonumber \\
q_{0|i} + q_{1|i} +q_{2|i} = 1 \\
q_{i|j}\rho_j = q_{j|i}\rho_i, \nonumber
\label{eq:Cons}
\end{eqnarray}
where $i,j = 0,1,2; i\ne j$; we suppress time dependence for simplicity.  These constraints leave only 5 of the 12 total variables are independent. Since species 2 is the invader, we track ${\rho_1,\rho_2, q_{2|2},q_{1|2},q_{1|1}}$. Note that the dynamics of ${q_{1|2}}$, the conditional density of a resident given an invader at a neighboring site, can depend on a novel-weapons effect when $\theta_2 > 0$.

The dynamics of the global densities, like the mean field approximation, account introduction to empty sites, local propagation, background mortality, and death due to interference competition. For the resident,
\begin{equation}\label{eq:GLDen:1}
\dot{\rho}_1 = \beta\rho_0 + \rho_1\alpha_1 q_{0|1} - \rho_1 \mu - \theta_2 q_{2|1} \rho_1\\
\end{equation}
Compared to the mean-field model, both local propagation and mortality due to interference depend on local, rather than global, densities.  Hence the PA models both site preemption and  interference as effects of spatially clustered growth.

To model the dynamics of a local density, we first write the dynamics of a doublet $\rho_{ii}$, a global density, where
$\rho_{ii} = \rho_i q_{i|i}$.  The PA's doublet dynamics ($d\rho_{ii}/dt$; see Appendix B) introduces the triplets ${q_{2|02}}$ and ${q_{1|22}}$. Ordinary PA assumes that neighbors of neighboring sites are weakly correlated, and lets ${q_{2|02} = q_{2|0}}$ and ${q_{1|22} = q_{1|2}}$. The approximation closes the system of equations, permitting us to write an invasion criterion \citep{Iwasa_1998}.

For the invasion analysis, assume that an introduction of species 2 has occurred.  Invasion either succeeds or fails before the next introduction event occurs (since $\beta \ll \mu < \alpha_i$).  Successful invasion requires that the invader have a positive growth rate when rare; $\dot{\rho_2} > 0$ as $\rho_2\rightarrow 0$.  From Appendix B, this condition yields the invasion criterion:
\begin{equation}\label{eq:InvCond}
1 - q_{2|2} - q_{1|2} > \frac{\mu + \theta_1 q_{1|2}}{\alpha_2}.
\end{equation}
The left side represents the frequency of open sites neighboring  an invader, $q_{0|2}$.  The right side represents the ratio of death rate to birth rate for the invader. Successful invasion, then, requires that the invader colonize a neighboring, empty site before it dies.  The death rate sums background mortality and averaged mortality from the resistance zone about a resident neighboring the invader.  The seeming simplicity of the invasion criterion masks an important biological difference between the MF criterion and the neighborhood-scale condition for invasion.  In the PA condition for successful invasion, the local density $q_{1|2}$, hence the local density of open sites $q_{0|2}$, depends on $\theta_2$, invader interference (see Appendix B).  That is, the neighborhood-scale invasion criterion reveals a role for novel weapons, contrasting to the MF result.  Consequently, an interfering invader might invade a resident species despite the resident having the greater propagation rate.  The right side of the criterion, of course, include the resistance-zone effect of interference by the resident.

To compare the PA invasion criterion to both the MF and simulation of the full spatial model (see below), we evaluated Inequality (\ref{eq:InvCond}) numerically.  We constructed pairwise invasion plots for the three cases defined above: only the invader interferes (novel weapons), only the resident interferes (resistance zones) and both species exercise interference competition.  Whenever a species was an interference competitor, we invoked the propagation-interference constraint.

\begin{figure}[b]
\centering
\vspace*{6.00truecm}
       \includegraphics{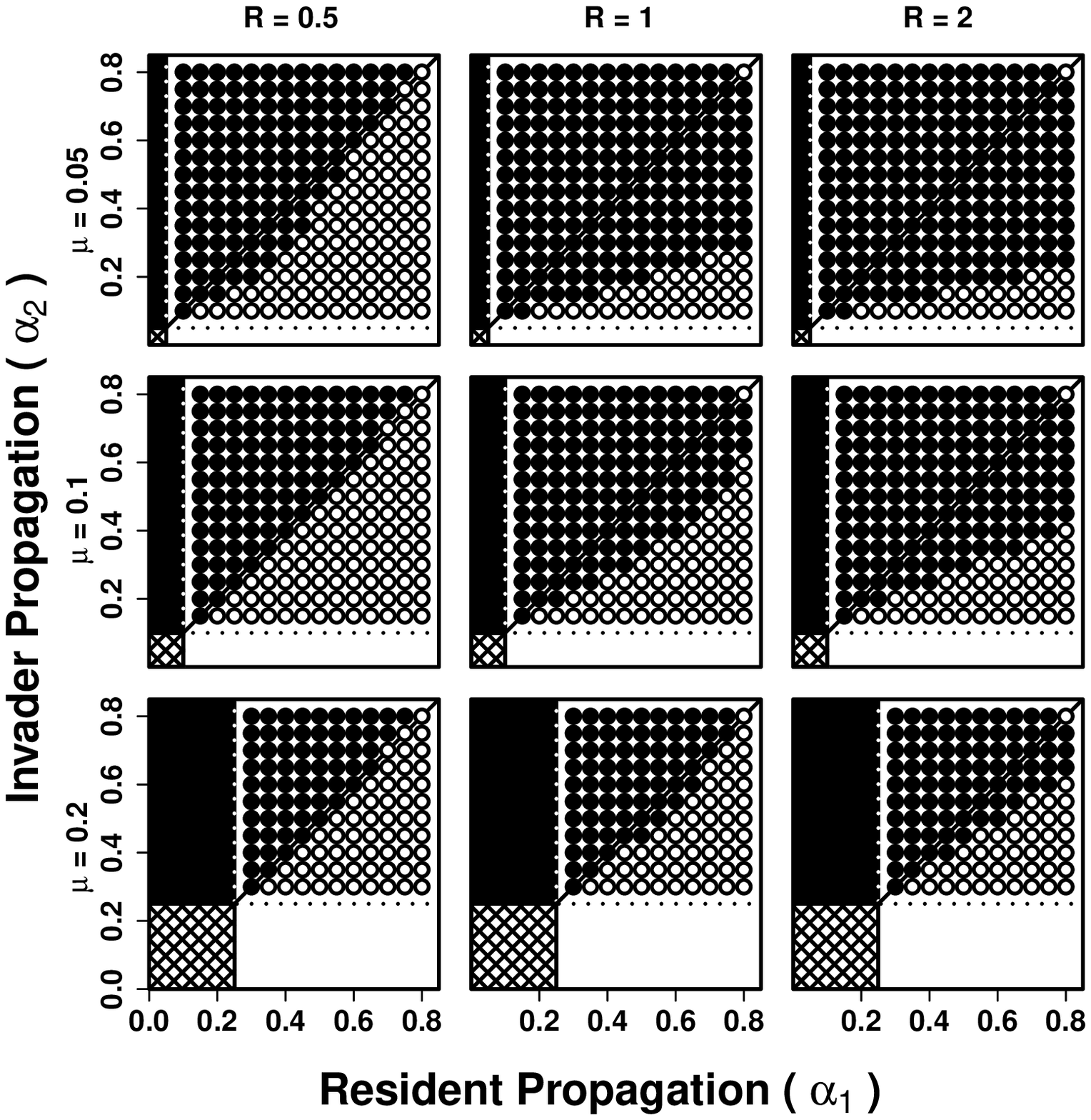}
\vspace*{4.0truecm}
\caption{
Novel weapons under pair approximation. Pairwise invasion plots generated by the PA invasion criterion; only the invader exercises interference.  Black: parameter combinations where invader's growth is positive when rare; invasion succeeds. White: resident repels invader; invasion fails.
}
\label{PAnovel}
\end{figure}

Figure \ref{PAnovel} shows pairwise invasion results when the invader, and not the resident, is the interference competitor.  That is, the invader obeys the propagation-interference constraint, and $\theta_1 = 0$ independently of the resident's propagation rate; we so isolate the novel weapons effect.  When interactions are local, interference competition can promote an invader's growth when rare.  When the propagation cost of interference is smaller ($R$ is larger) and $\mu$ is small, invaders with a propagation rate much less than the resident's rate succeed.  Invader interference proves advantageous since it opens sites neighboring invaders.  Hence interference makes space available where the invader's local density may match or exceed the resident's local density - despite the resident's greater global density.  Hence, novel weapons, in our model, enhances invasion when individuals compete at the neighborhood scale, but has no effect under homogeneous mixing.

\begin{figure}[b]
\centering
\vspace*{6.00truecm}
       \includegraphics{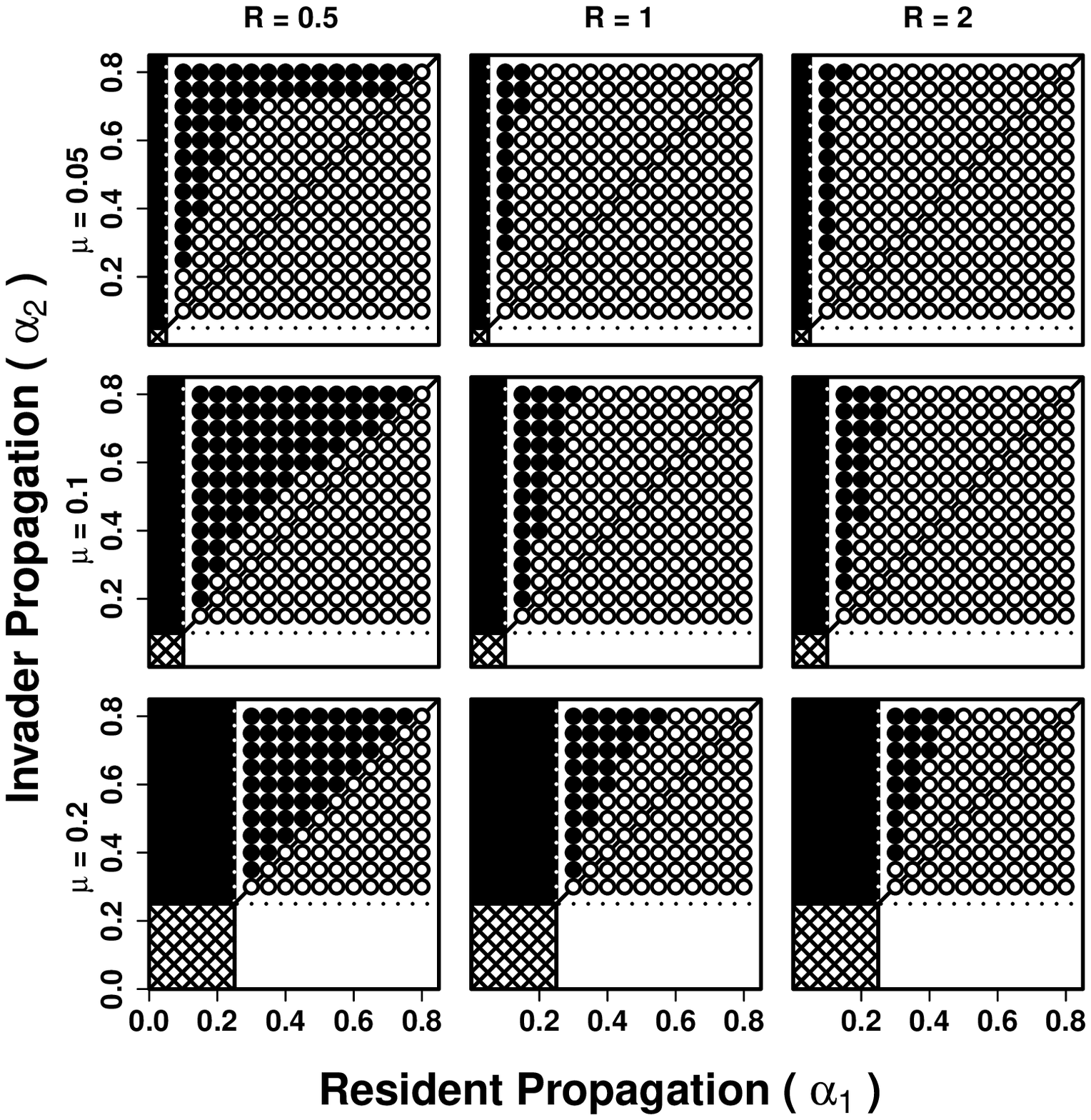}
\vspace*{4.0truecm}
\caption{
Resistance zones under pair approximation. Pairwise invasion plots generated by the PA invasion criterion; only the resident exercises interference.  Black: parameter combinations where invader's growth is positive when rare; invasion succeeds. White: resident repels invader; invasion fails.
}
\label{PAresist}
\end{figure}

Figure \ref{PAresist} shows pairwise invasion results when the resident, and not the invader, is the interference competitor.  That is, the resident obeys the propagation-interference constraint, and $\theta_2 = 0$ independently of the invader's propagation rate; we isolate the resistance zone effect.  A resident can repel an invader with a much higher propagation rate than its own through local interference.  Increasing $R$ now reduces the range of parameter combinations permitting invasion, and increasing $\mu$ promotes invader success - results symmetrically opposite to those for novel weapons.  When only the resident can interfere, a greater value for $R$ reduces the pleiotropic cost of interference.  Increasing $\mu$ opens more space for the invader; greater background mortality renders a given level of interference less effective as a competitive mechanism.

\begin{figure}[b]
\centering
\vspace*{6.00truecm}
       \includegraphics{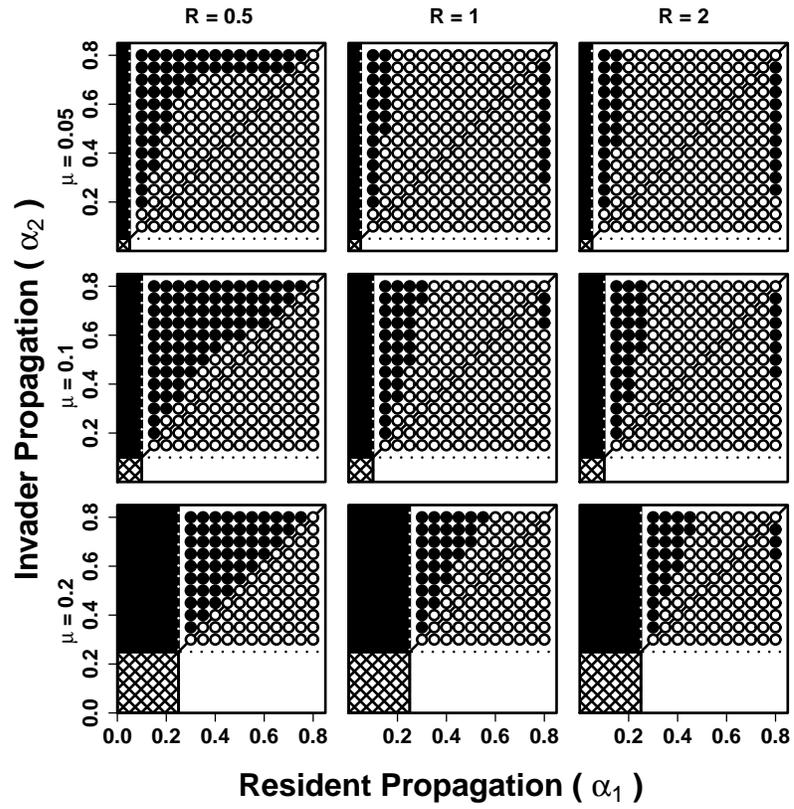}
\vspace*{4.0truecm}
\caption{
Pairwise invasion plots generated by the PA invasion criterion; both species exercise interference.  Black: parameter combinations where invader's growth is positive when rare; invasion succeeds. White: resident repels invader; invasion fails.
}
\label{PAboth}
\end{figure}

Figure \ref{PAboth} shows pairwise invasion results when both species exercise interference, and so both are constrained by the propagation-interference tradeoff.  Not surprisingly, the resident species, because of its initial high density, can employ interference to repel the invader over a much greater range of parameters than when both species lack interference (where greater propagation always wins).  The utility of interference, from the resident species' perspective, declines at small $R$, and at greater $\mu$; the invasion plot for $(R = 0.5, \mu = 0.2)$ matches the case where competition is strictly preemptive. In most of the plots substantial regions of bistability appear.  A species with intermediate levels of both propagation and interference can repel a broad range of propagation-interference combinations.  And, when rare, the intermediate combination is repelled by species within that broad range, as long as $R \geq 1$.

Mutual interference does not permit coexistence under PA (fig. \ref{PAboth}).  In general, the invasion plot resembles the case where only the resident can interfere.  But for sufficiently large $R$ and small enough $\mu$, a resident that invests solely in propagation, and not in interference, can be invaded by a species that mixes propagation and interference (rightmost columns of invasion plots).  The resident, with an initially high density and no interference capacity, is vulnerable to a clustered invader that can locally open sites \emph{via} interference competition.
\section{Simulation of individual-based model}
We simulated the individual-based spatial model with $L_x = L_y \equiv L = 256$, and $\beta$ = 0.001. We initiated simulations with the resident occupying each site.  The resident's density was allowed to decline to its single species equilibrium density, $\rho_1^*$, without any introduction.  Then, at time $t = 0$, introduction of both invader and resident individuals began. We tracked the global densities of each species $\rho_i(t)$, and recorded a successful invasion when the invader reduced the global density of the resident species to $\rho_1^*/2$ within 20,000 MCSS time steps.  These simulations envision growth of initially small invader clusters at the initiation of ecological invasion \citep{Korniss_JTB2005,OMalleyTPB_2006,Allstadt_2007}.

We can attribute some differences between results of the individual-based spatial model and its pair approximation to the difference in degree of spatial clustering.  PA, by construction, truncates correlation length at a nearest neighbors.  The simulation model permits development of greater correlation lengths (invader clusters can grow large from a single site).  Cluster size impacts the relative frequency at which individuals experience local self-regulation versus interspecific competition \citep{OMalley_2010}.  These frequencies, in turn, affect invasion success and global population densities.

\begin{figure}[b]
\centering
\vspace*{6.00truecm}
       \includegraphics{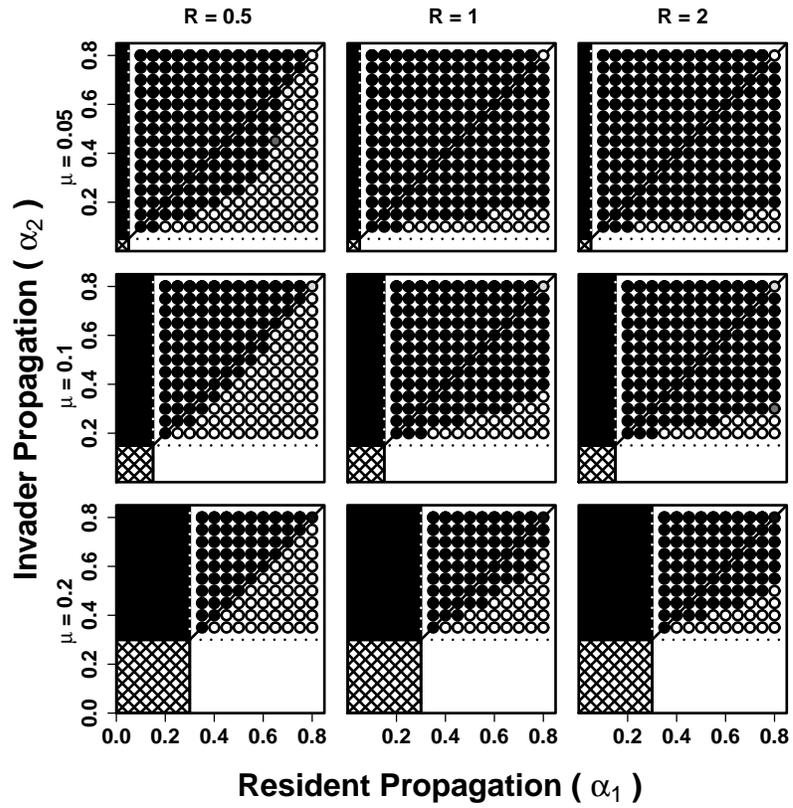}
\vspace*{4.0truecm}
\caption{
Novel weapons in the simulation model of invader clusters.  Each point represents combined results of 20 simulations. Black: parameter combinations where invader's growth is positive when rare; invasion succeeds. White: resident repels invader; invasion fails.  Stochasticity led to mixed results for a few parameter combinations, which are shaded gray according to the proportion of successful invasion.
}
\label{SimNovel}
\end{figure}

We simulated invasion from rare introduction for the three cases examined above: novel weapons, resistance zones, and the case where both species interfere competitively.  Figure \ref{SimNovel} shows pairwise invasion results when the invader, and not the resident, interferes competitively in the individual-based spatial model.  The invader obeys the propagation-interference constraint, and $\theta_1 = 0$ independently of the resident's propagation rate, isolating the novel weapons effect.  Invader interference promotes successful invasion more frequently in simulation than under pair approximation.  That is, we note a novel-weapons effect (absent in the MF).  Invader clustering increases the advantage of local interference competition against a resident lacking interference.  Even when the propagation-cost of interference is steep ($R < 1$) an invader may advance and exclude a resident with a greater propagation rate.  When interference is less costly ($R > 1$) most invaders succeed, except those with a very low propagation rate.  PA, then gives a reasonable, but understated, prediction of the impact of novel weapons in the detailed spatial model.

\begin{figure}[b]
\centering
\vspace*{6.00truecm}
       \includegraphics{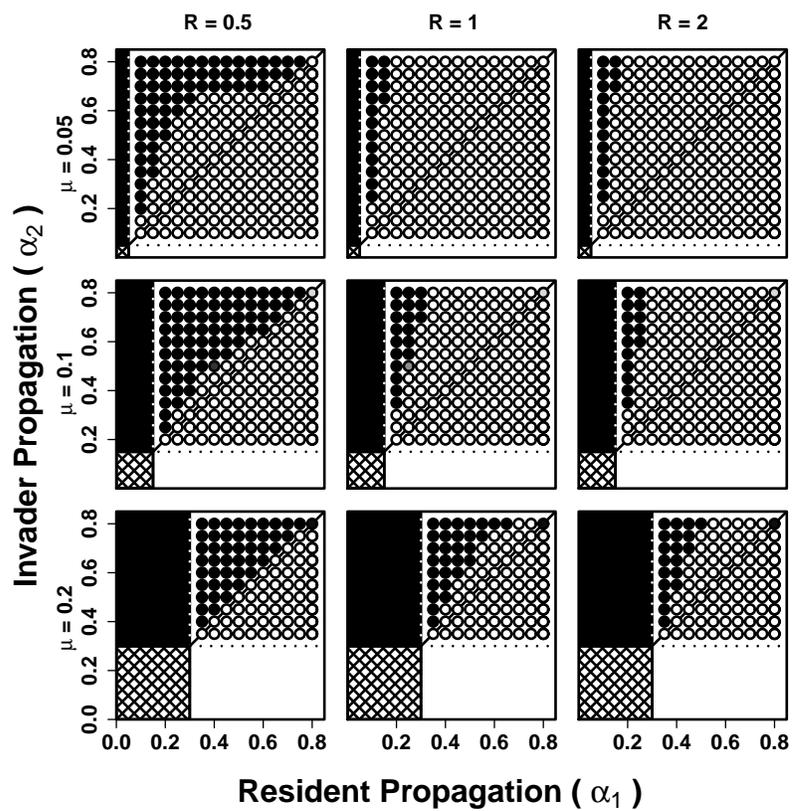}
\vspace*{4.0truecm}
\caption{
Resistance zones in the simulation model of invader clusters.  Each point represents combined results of 20 simulations. Black: parameter combinations where invader's growth is positive when rare; invasion succeeds. White: resident repels invader; invasion fails.  Gray: as described above.
}
\label{SimResist}
\end{figure}

Figure \ref{SimResist} shows pairwise invasion results when the resident, and not the invader, interferes competitively in the individual-based model.  The resident obeys the propagation-interference constraint, and $\theta_2 = 0$ independently of the invader's propagation rate, isolating the resistance zone effect.  Interference allows a resident to repel invasion by rare species with much greater rates of local propagation.  For the resistance-zone effect, PA predicts the results of the individual-based model very accurately across all parameter combinations we simulated.

\begin{figure}[b]
\centering
\vspace*{6.00truecm}
       \includegraphics{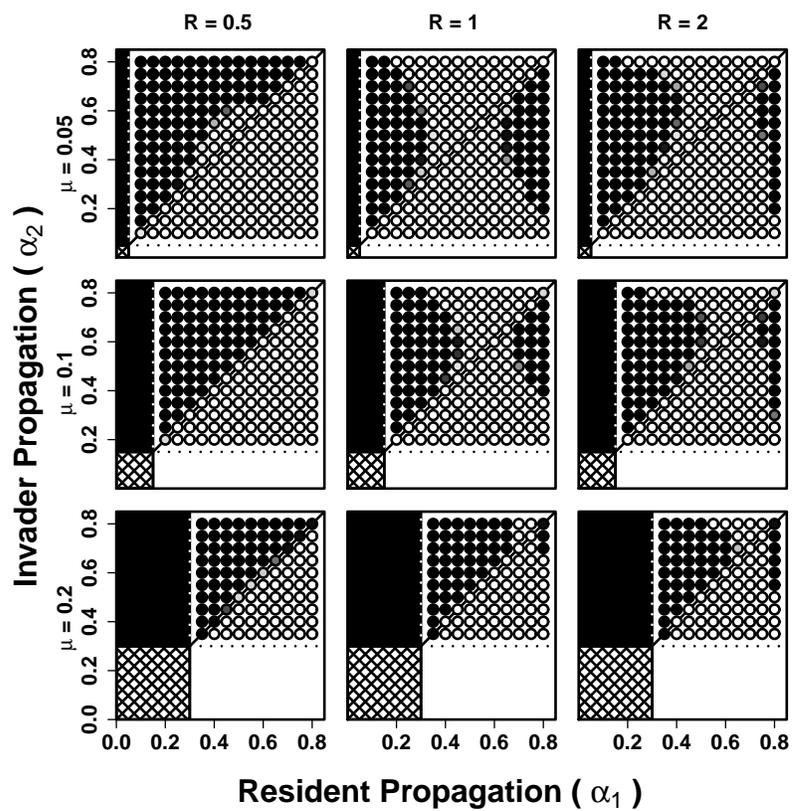}
\vspace*{4.0truecm}
\caption{
Simulation model: invader clusters when both species interfere.  Each point represents combined results of 20 simulations. Black: parameter combinations where invader's growth is positive when rare; invasion succeeds. White: resident repels invader; invasion fails.  Gray: as described above.
}
\label{SimBoth}
\end{figure}

The simulation model's pairwise invasion results for the case where both species exercise interference competition appear in Figure \ref{SimBoth}.  When the cost of interference is high ($R = 0.5$) the results are very close to those for preemptive competition only.  That is, the species with the greater rate of propagation wins in most cases.  Compared to PA, invader clustering in the simulation model produces more cases of successful invasion at low background mortality.  At lesser costs of interference $(R = 1, 2)$ residents mixing intermediate levels of propagation and interference repel most invading species, particularly when background mortality is not too large.  Residents that invest either too much in interference (low $\alpha_1$) or too little are susceptible to invasion by species with a more balanced mix of propagation and interference.

In a separate exercise, we released both species from the propagation-interference constraint and searched the parameter space for competitive coexistence. We set $\mu = 0.1$, $\beta = 0.001$, and let $L = 64$ to save computation time. The simulations ran for 100,000 MCSS, and the threshold for coexistence was $\rho_i > 0.05; i = 1,2$, well above the expected density due to introduction alone. The search yielded no evidence of competitive coexistence, as our analytical models predict.

\section{Discussion}
In the novel weapons scenario an invader gains competitive superiority over a resident through interference.  Applications focus on allelopathic interference \citep{Callaway_2000}.  If species mix homogeneously, we find that a rare invader gains no advantage through interference competition \citep{Case_1974}.  But if competitive interactions occur at the local scale, an invader can employ interference to advance against a resident species with a greater rate of local propagation (figs. \ref{PAnovel} and \ref{SimNovel}).  Invaders are rare globally but locally clustered in our spatial models.  Invaders can use interference to reduce resident density at the perimeter of these  clusters, where local invader density is sufficiently high to compete effectively for open sites \citep{Korniss_JTB2005}.

In the resistance zone scenario, the resident holds competitive superiority over an invader through interference.  A resident's competitive interference has a strong effect under homogeneous mixing; few invaders advance successfully unless interference is costly and background mortality is high (fig. \ref{MF}).  Under pair approximation, interference by the resident restricts invasion in two cases: when the resident, but not the invader, interferes, and when both species exercise interference competition.  The resistance zone effect also appears in simulation of the individual-based model.  But when both species can interfere, successful invasion is far more common in the simulation.  The difference between the full spatial model and its pair approximation lies in the impact of invader clusters larger than neighborhood size (hence, longer correlation distances) generated by the individual-base model.  Clustering increases an invader's benefit from interference against an interfering resident.

When both species exhibit interference, the pairwise invasion plots indicate regions of bistability (if only one species interferes, one cannot meaningfully reflect invasion plots about the diagonal).  We found large regions of bistability under homogeneous mixing.  Bistability declined as spatial structure increased \citep{Chao_1981}, and nearly disappeared in the simulation model (fig. \ref{SimBoth}).  Given a sufficiently long time, we would anticipate that the spatial process on a large, but finite, lattice would result in competitive exclusion.  The time required, however, may exceed population-dynamic scales.

In a three-species spatial model, where a non-interfering species interacted with both a weak interferer and a strong interferer, \citet{Durrett_1997} found cyclic coexistence. Our models for two-species competition, by construction, do not admit coexistence.  In fact, when we assumed that both species exercise interference competition, and when increased interference reduces propagation, certain parameter sets suggested that a single combination of propagation and interference might repel all other feasible combinations.  Suppose we start the resident's propagation rate, $\alpha_1$, at minimal (maximal) levels.  Then larger (smaller) propagation rates successively invade and exclude the resident until pairwise competition leaves a resident species that can exclude all others.

\section*{Acknowledgments}
The National Science Foundation supported this research through Grants DEB-0918392 and DEB-0918413.  We thank J. Newman, G. Robinson, I.-N. Wang and A.C. Gorski for comments.  D. Yoakam kept us energized.
\newpage
\newpage
\newpage
\newpage
\newpage
\newpage
\newpage
\newpage

\appendix

\section{Mean-field stability}
The Jacobian for the spatially homogeneous MF dynamics is:
\begin{equation}
J =
\left[
  \begin{array}{cc}
    \alpha_1 - \mu - 2\alpha_1 {\rho_1}^* - {\rho_2}^* (\alpha_1 + \theta_2) & - {\rho_1}^*(\alpha_1 + \theta_2) \\
    - {\rho_2}^* (\alpha_2 + \theta_1) & \alpha_2 - \mu - 2\alpha_2 {\rho_2}^* - {\rho_1}^* (\alpha_2 + \theta_1) \\
  \end{array}
\right]
\label{Jacob}
\end{equation}
Evaluated at $E1$, mutual extinction, the Jacobian becomes:
\begin{equation}
J(E1) =
\left[
  \begin{array}{cc}
    \alpha_1 - \mu  & 0 \\
    0 & \alpha_2 - \mu  \\
  \end{array}
\right]
\end{equation}
Since $\alpha_i > \mu$ for $i = 1,~2$, by assumption, mutual extinction is unstable.

At $E2$ the resident excludes the invader.  The associated Jacobian yields:
\begin{equation}
J(E2) =
\left[
  \begin{array}{cc}
    \alpha_1 - \mu - 2\alpha_1 (1 - \frac{\mu}{\alpha_1}) & - (1 - \frac{\mu}{\alpha_1})(\alpha_1 + \theta_2) \\
    0 & \alpha_2 - \mu - (1 - \frac{\mu}{\alpha_1}) (\alpha_2 + \theta_1) \\
  \end{array}
\right]
\label{JacobE2}
\end{equation}
The two eigenvalues are:
\begin{eqnarray}
\lambda_1(E2) = \mu (\frac{\alpha_2}{\alpha_1} - 1) + \theta_1 (\frac{\mu}{\alpha_1} - 1)\\
\lambda_2(E2) = \mu - \alpha_1 < 0
\label{EigenE2}
\end{eqnarray}
Local stability requires $\lambda_1(E2) < 0$.

For species 2 to advance when rare, $\lambda_1(E2)$ must be positive (implying that extinction of
species 2 is unstable).  Note that if $\alpha_2 > \alpha_1$ , then
$(\frac{\alpha_2}{\alpha_1} - 1) > 0$, a condition promoting invasion by species 2 when species 1 rests its single-species equilibrium $E2$.  When $\theta_1 > 0$, species 2 invades \emph{iff}:
\begin{equation}
\alpha_2 - \alpha_1 > \theta_1 (\frac{\alpha_1}{\mu} - 1) > 0
\label{2invades_app}
\end{equation}

Next, consider $E3$, where species 2 competitively excludes species 1.  Evaluating the Jacobian,
we have:
\begin{equation}
J(E3) =
\left[
  \begin{array}{cc}
      \alpha_1 - \mu - (1 - \frac{\mu}{\alpha_2}) (\alpha_1 + \theta_2) & 0\\
    - (1 - \frac{\mu}{\alpha_2})(\alpha_2 + \theta_1)  & \alpha_2 - \mu - 2\alpha_2 (1 - \frac{\mu}{\alpha_2}) \\
  \end{array}
\right]
\label{JacobE3}
\end{equation}
The symmetry between $J(E2)$ and $J(E3)$ extends to the eigenvalues.  From Eq~\ref{JacobE3}, we obtain:
\begin{eqnarray}
\lambda_1(E3) = \mu (\frac{\alpha_1}{\alpha_2} - 1) + \theta_2 (\frac{\mu}{\alpha_2} - 1)\\
\lambda_2(E3) = \mu - \alpha_2 < 0
\label{EigenE3}
\end{eqnarray}
Species 2 cannot exclude species 1 \emph{iff}
\begin{equation}
\alpha_1 - \alpha_2 > \theta_2 (\frac{\alpha_2}{\mu} - 1) > 0
\label{1invades}
\end{equation}
Clearly, reversing subscripts in Expression~(\ref{2invades}) yields Expression~(\ref{1invades}).  More importantly, we see that if Inequality (\ref{2invades}) holds, then $\lambda_1(E3) < 0$.  Hence, if species 2 can invade species 1, $E3$ is the only stable equilibrium, and  species 2 advances to exclude species 1 competitively.

Now consider bistability.  When common, species 1 repels invasion by species 2 if $\lambda_1(E2) < 0$.  When species 2 is common, it repels invasion by species 1 if $\lambda_1(E3) < 0$.  Both conditions hold, and the dynamics is bistable, \emph{iff}:
\begin{equation}
\theta_1 (1 - \frac{\alpha_1}{\mu}) < \alpha_1 - \alpha_2~~\textrm{and~}~
\theta_2 (\frac{\alpha_2}{\mu} - 1) > \alpha_1 - \alpha_2
\label{bistable_app}
\end{equation}
From our analysis of the single-species equilibria, the two species can coexist \emph{iff} $\lambda_1(E2) > 0$ and $\lambda_1(E3) > 0$.  The first inequality implies that when species 1 rests at its single-species equilibrium, species 2 can advance from rarity.  The second inequality reverses roles of common and rare. In the text we show that the mean-field dynamics does not admit competitive coexistence.

\section{Pair-correlation dynamics}

\def\qzt{1 - q_{2|2} - q_{1|2}}
\def\pz{1-\rho_1-\rho_2}
\def\qzo{1-q_{1|1}-q_{1|2}\frac{\rho_2}{\rho_1}}
\def\qto{q_{1|2}\frac{\rho_2}{\rho_1}}
\def\qtz{\bigl(\frac{\rho_2}{1-\rho_1 -\rho_2}\bigr)(1- q_{2|2} - q_{1|2})}
\def\qoz{\frac{\rho_1}{1-\rho_1 -\rho_2}\bigl(\qzo\bigr)}
\def\deldel{\frac{\delta - 1}{\delta}}
\def\onedel{\frac{1}{\delta}}
We develop our pair approximation (PA) by modifying methods described by \citet{Iwasa_1998}.  We write a dynamics for five state variables: ${\rho_1,\rho_2, q_{2|2},q_{1|2},q_{1|1}}$.  The first two are global densities, and the other three are local densities.  Invoking constraints listed in the text, Expression (\ref{eq:Cons}), we express the remaining PA variables in terms of the five state variables.  In particular:
\begin{eqnarray}
\rho_0 = \pz ~~~~~~~~~~~~~~~~~~~~~~ q_{1|0} = \qoz \nonumber \\
q_{2|0} = \qtz ~~~~~~~~~~~~~~~~~~~ q_{0|2} = \qzt \nonumber \\
q_{0|1} = \qzo ~~~~~~~~~~~~~~~~~~~~~~~~~~~~~~~~~~~~~~~~~~~~~ q_{2|1} = \qto \nonumber \\
\label{eq:Conv}
\end{eqnarray}
We omit $q_{0|0}$, also easily calculated, since it does not appear in the analysis.

To begin, we rewrite the dynamics of the resident's global density, Eq. (\ref{eq:GLDen:1}), in terms of our five state variables.  Doing the same for the invader's global density yields:
\begin{eqnarray}\label{eq:GLDen}
\dot{\rho}_1 = \beta(\pz) + \rho_1 \bigl[\alpha_1(\qzo)- \mu - \theta_2\qto \bigr] \\
\dot{\rho}_2 = \beta(\pz) + \rho_2 \bigl[\alpha_2(\qzt)- \mu - \theta_1 q_{1|2} \bigr]
\end{eqnarray}

To write the dynamics of the three conditional densities, we first require the dynamics of a corresponding doublet.  By definition, ${q_{j|i} = \frac{\rho_{ij}}{\rho_i}}$, where ${\rho_{ij}}$ is the unordered doublet density. The doublet $\rho_{ij}$ has dynamics:
\begin{equation} \label{eq:q22dot:1}
\dot{q}_{j|i} = \frac{1}{\rho_i}\dot{\rho}_{ij} - \frac{{q}_{j|i}}{\rho_i}\dot{\rho}_i
\end{equation}
Applying the recipe to ${q_{2|2}}$, we have:
\begin{equation} \label{eq:rho22:1}
\dot{\rho}_{22} = 2\rho_{02}\beta + 2\rho_{02}\alpha_2 \Bigl[\onedel + \deldel q_{2|02} \Bigr] - 2 \rho_{22} \Bigl[\mu + \deldel \theta_1 q_{1|22}\Bigr]
\end{equation}
The first two terms represent generation of new invader pairs through introduction and birth, and the third term represents the loss of invader pairs due to background mortality and interference competition. This equation introduces triplets ${q_{2|02}}$ and ${q_{1|22}}$ into the dynamics. Ordinary PA takes neighbors of neighboring sites as weakly correlated, and so we assume that $q_{2|02} = q_{2|0}$ and $q_{1|22} = q_{1|2}$. The resulting closure of the equations allows the analysis without including the 27 types of triplets, or any higher order spatial correlations.

Using the closure assumption and converting terms with equation (\ref{eq:Conv}), equation (\ref{eq:rho22:1}) becomes
\begin{eqnarray} \label{eq:rho22:2}
\dot{\rho}_{22} =  2\rho_2(\qzt)\Bigl[\beta + \alpha_2 \bigl(\onedel + \frac{\rho_2(\delta-1)}{\delta(\pz)}(\qzt)\bigr)\Bigr]   \nonumber \\
-2\rho_2 q_{2|2}\Bigl[\mu + \deldel \theta_1 q_{1|2}\Bigr]~~~~~~~~~~~~~~~~~~~~
\end{eqnarray}
Substituting Eqq. (\ref {eq:GLDen:1}) and (\ref{eq:rho22:2}) into Eq. (\ref{eq:q22dot:1}) yields:
\begin{eqnarray} \label{eq:q22dot:2}
\dot{q}_{2|2} = 2(\qzt)\Bigl[\beta + \frac{\alpha_2}{\delta} \bigl(1 + \frac{\rho_2(\delta-1)}{\pz}(\qzt)\bigr)\Bigr] \nonumber \\ - q_{2|2} \Bigl[\frac{\beta}{\rho_2}(\pz) + \alpha_2(\qzt) - \mu - \theta_1 q_{1|2} \Bigr] \nonumber \\ -2 q_{2|2}\Bigl[\mu + \deldel \theta_1 q_{1|2}\Bigr]~~~~~~~~~~~~~~~
\end{eqnarray}
Similarly,
\begin{eqnarray} \label{eq:q11dot}
\dot{q}_{1|1} = 2(\qzo)\Bigl[\beta + \frac{\alpha_1}{\delta} \bigl(1 + \frac{\rho_1(\delta-1)}{\pz}(\qzo)\bigr)\Bigr] \nonumber \\
- q_{1|1} \Bigl[\frac{\beta}{\rho_1}(\pz)+ \alpha_1(\qzo) - \mu - \theta_2 \qto \Bigr] \nonumber \\
-2 q_{1|1}\Bigl[\mu + \deldel \theta_2 \qto\Bigr]~~~~~~~~~~~~~~~~~
\end{eqnarray}
${\dot{\rho}_{12}}$ is slightly more complicated.  There are two ways to make a (1, 2) pair, and each member of a (1, 2) pair interferes with the other. Proceeding:
\begin{eqnarray} \label{eq:rho12}
\dot{\rho}_{12} = \rho_2 (\qzt) \Bigl[\beta + \frac{\alpha_1 \rho_1(\delta-1)}{\delta(\pz)}(\qzo) \Bigr] \nonumber \\
+ \rho_1 (\qzo) \Bigl[\beta + \frac{\alpha_2 \rho_2 (\delta-1)}{\delta(\pz)}(\qzt) \Bigr]
\nonumber \\
- \rho_2 q_{1|2} \Bigl[2 \mu + \frac{\theta_1}{\delta}\bigl(1 + (\delta-1) q_{1|2}\bigr) +
\frac{\theta_2}{\delta}\bigl(1 + (\delta-1) \qto \bigr) \Bigr]
\end{eqnarray}
Then, following substitution:
\begin{eqnarray} \label{eq:q21dot}
\dot{q}_{1|2} = (\qzt)\Bigl[\beta + \frac{\alpha_1 \rho_1(\delta-1)}{\delta(\pz)}(\qzo)\Bigr]\nonumber \\
+ \rho_1 (\qzo) \Bigl[\frac{\beta}{\rho_2} + \frac{\alpha_2(\delta-1)}{\delta(\pz)}(\qzt)  \Bigr] \nonumber \\
- q_{1|2} \Bigl[ \mu + \frac{\theta_1}{\delta}\bigl(1 + (\delta-1) q_{1|2}\bigr) +
\frac{\theta_2}{\delta}\bigl(1 + (\delta-1) \qto \bigr) \Bigr]
\nonumber \\
- q_{1|2} \Bigl[ \frac{\beta(\pz)}{\rho_2} + \alpha_2(\qzt)
-q_{1|2} \theta_1 \Big]
\end{eqnarray}
Equations (\ref{eq:GLDen}), (B.3), (\ref{eq:q22dot:2}), (\ref{eq:q11dot}), and (\ref{eq:q21dot}) constitute the pair-approximation dynamics.

To analyze invasion, we introduce species 2 at near-zero density.  Invasion succeeds or fails before the next introduction event occurs (since $\beta \ll \mu_i < \alpha_i$). Taking $\beta = 0$, the PA dynamics becomes:
\begin{eqnarray}
\dot{\rho}_1 = \rho_1 \bigl[\alpha_1(\qzo)- \mu - \theta_2\qto \bigr]~~~~~ \label{eq:InvDy1} \\
\dot{\rho}_2 = \rho_2 \bigl[\alpha_2(\qzt)- \mu - \theta_1 q_{1|2} \bigr]~~~~~
\label{eq:InvDy2}\\
\dot{q}_{2|2} = 2\alpha_2(\qzt)\Bigl[\onedel + \frac{\rho_2(\delta-1)}{\delta(\pz)}(\qzt)\Bigr]~~~~~ \label{eq:InvDy3}\\ \quad - q_{2|2} \Bigl[\alpha_2(\qzt) + \mu + \theta_1 q_{1|2} \bigl(\frac{2(\delta-1)}{\delta} - 1 \bigr) \Bigr] ~~~~~ \nonumber\\
\dot{q}_{1|1} = 2\alpha_1(\qzo)\Bigl[\onedel + \frac{\rho_1(\delta-1)}{\delta(\pz)}(\qzo)\Bigr] ~~~~~ \label{eq:InvDy4}\\ \quad - q_{1|1} \Bigl[\alpha_1(\qzo) + \mu + \theta_2 \qto \bigl(\frac{2(\delta-1)}{\delta} - 1 \bigr) \Bigr] ~~~~~ \nonumber\\
\dot{q}_{1|2} = (\qzt)(\qzo)\bigl( \frac{\rho_1(\delta-1)}{\delta(\pz)}\bigr) (\alpha_1+\alpha_2) ~~~~~ \label{eq:InvDy5}\\  \quad -q_{1|2} \Bigl[\mu + \alpha_2 (\qzt)\Bigr]~~~~~ \nonumber \\  -q_{1|2} \Bigl[
\theta_1\bigl(\onedel + \deldel q_{1|2} -  q_{1|2} \bigr) +
\theta_2\bigl(\onedel + \deldel \qto \bigr) \Bigr]~~~~~
\nonumber
\end{eqnarray}
The next step of the invasion analysis addresses the frequency of open sites when the invader is rare. Since the invader is rare, species 2 has no effect on either the resident's equilibrium global density ($\rho_1^*$) or the equilibrium frequency of paired residents ($q_{1|1}^*)$. Setting $\rho_2 = 0$ in Eqq. (\ref{eq:InvDy1}) and (\ref{eq:InvDy4}) yields:
\begin{equation}\label{eq:rho1star}
\rho_1^* = \frac{\delta-1-\frac{\mu}{\alpha_1}\delta}{\delta-1-\frac{\mu}{\alpha_1}}
\end{equation}
and
\begin{equation}\label{eq:q11star}
q_{1|1}^* = 1 - \frac{\mu}{\alpha_1}.
\end{equation}
Now we use these results to find the other conditional probabilities.
Let $x = q_{2|2}$, $y = q_{1|2}$, and $w = q_{0|2} = 1 - x - y$. Then:
\begin{eqnarray}\label{eq:xydot}
\dot{x} = \frac{2\alpha_2 w}{\delta} + xy\theta_1 \Bigl[ 1 - \frac{2(\delta-1)}{\delta}\Bigr] - \alpha_2 w x - \mu x \nonumber\\
\dot{y} = \bigl(1-\frac{1}{\delta}-\frac{\mu}{\alpha_1} \bigr)\bigl(\alpha_1 + \alpha_2\bigr) w + y \Bigl[\alpha_2 w -\mu - \frac{\theta_1}{\delta} - \frac{\theta_2}{\delta} \Bigr]  \\  \qquad + y^2 \theta_1\Bigl[1-\frac{\delta-1}{\delta} \Bigl] ~~~~~~~~~~~~~~~~~~~~ \nonumber
\end{eqnarray}
Solving for the equilibria, we have:
\begin{eqnarray}\label{eq:xystar}
x^* = \frac{2\alpha_2 w}{\alpha_2\delta w + \mu \delta + y^* \theta_1 (\delta-2)}~~ \nonumber \\
\\
y^* = \frac{w(\alpha_1 +\alpha_2) (\delta - 1 - \delta \frac{\mu}{\alpha_1})}{\delta \left[\mu + \alpha_2 w \right] + \theta_1 - y^* \theta_1 + \theta_2}
\end{eqnarray}
Given the interdependence of $x^*$ and $y^*$, we evaluated $(\dot{\rho_2})_{\rho_2 = 0} > 0$ numerically.  Using Eq. (\ref{eq:InvDy2}), we obtain the invasion criterion, Eq. (\ref{eq:InvCond}) in the text.  The importance of deriving $x^*$ and $y^*$ lies in demonstrating that at the neighborhood scale, both species' level of interference ($\theta_1$ and $\theta_2$) affects the likelihood that species 2 can invade the resident species 1.  That is, the invasion criterion defined  by  neighborhood-scale correlations depends not only on the resident's level of interference $\theta_1$ (resistance zone), but on the invader's level of interference $\theta_2$ (novel weapons).

\bibliographystyle{elsarticle-harv}




\end{document}